\documentclass[10pt]{iopart}

\usepackage{iopams}  
\usepackage{graphicx}
\usepackage[breaklinks=true,colorlinks=true,linkcolor=blue,urlcolor=blue,citecolor=blue]{hyperref}
\usepackage{epstopdf}
\usepackage{caption}

\begin{document}

\title[Temperature and composition dependence of the band gaps of Ga$_{1-x}$In$_x$N alloy: a first-principles study based on the virtual crystal approximation]{Temperature and composition dependence of the band gaps of Ga$_{1-x}$In$_x$N alloy: a first-principles study based on the virtual crystal approximation}

\author{Jian Liu}
\address{Department of Physics and Astronomy, Stony Brook University, Stony Brook, NY 11794-3800, United States.}
\eads{\mailto{Jian.Liu@stonybrook.edu}}

\begin{abstract}
We report on the structural, electronic and vibrational properties of the Ga$_{1-x}$In$_x$N alloy using virtual crystal approximation (VCA) from first-principles. A band gap bowing parameter of 3.85 eV is obtained with the TB09 functional. Phonon density of states shifts to lower frequency as the In content is increased. However, VCA ignores disorder effect and is therefore unable to describe the broadening of the phonon spectra upon alloying. The role of electron-phonon interaction in the temperature dependence of the band gap is also studied for GaN, InN and their alloy Ga$_{1-x}$In$_x$N. The calculated zero-point motion renormalization and the fitted Varshni parameter over the entire composition range are discussed.
\begin{description}
\item[PACS numbers]
\end{description}
\end{abstract}

\maketitle

\ioptwocol

\section{Introduction}
The Ga$_{1-x}$In$_x$N alloy has attracted great interest due to its technological importance in optoelectronics and electronics. Direct generation of hydrogen by splitting water using solar energy has also been reported for the Ga$_{1-x}$In$_x$N alloy\cite{GaInN-photo}. The band gap of the Ga$_{1-x}$In$_x$N alloy can be tuned to cover nearly the entire solar spectrum, from 0.8 eV (InN) to 3.5 eV (GaN), by varying the composition $x$. Many basic properties of the Ga$_{1-x}$In$_x$N alloy are not well known due to the difﬁculty in growing high quality samples. The composition-dependence of the band gap, also known as the band gap bowing, has been extensively studied in recent years. However, consensus on the band gap bowing of the Ga$_{1-x}$In$_x$N alloy has not been reached. A bowing parameter $b$ of 1.43 eV is reported from an early measurement\cite{GaInN-exp-1.43}, while more recently a considerably larger value of 2.8 eV is experimentally observed\cite{GaInN-exp-2.8}. Results from first-principles calculations range from 1.1 eV (HSE06)\cite{GaInN-HSE06}, 1.3 eV (LDA-1/2)\cite{GaInN-LDA12} and 1.5 eV (mBJ)\cite{GaInN-mBJ} to 2.1 eV (LDA+C)\cite{GaInN-LDA+C}, depending on the exchange-correlation functional used. The enhancing effect of Indium clustering on the band gap bowing has also been examined\cite{GaInN-LDA+C}\cite{GaInN-CE}. Previous first-principles calculations rely on either one reasonably sized special quansirandom structure (SQS) supercell or a statistical ensemble of clusters to account for the compositional disorder. The virtual crystal approximation (VCA), on the other hand, has the main advantage of simplicity in modelling the disordered alloy. VCA has demonstrated good accuracy for semiconductor alloys and ferroelectric perovskite solid solutions\cite{ferro-VCA-1}\cite{ferro-VCA-2}. It offers an efﬁcient way of studying quantities which are computationally more demanding such as vibrational properties and electron-phonon interactions. Indeed it is found that the inclusion of lattice vibrations, which is commonly neglected due to its prohibitive computational cost, could reduce the order-disorder transition temperature by $\sim$30\%\cite{SQS-wz}.
\newline
In the recent past, the temperature dependence of the band gaps $E_g(T)$ of semiconductors have attracted increasing attentions\cite{EPI-LCAO}\cite{EPI-Wannier}\cite{EPI-BSE}\cite{EPI-C}\cite{EPI-optSi}\cite{EPI-Giustino}\cite{EPI-adiabatic}\cite{EPI-Ponce}. Experimentally, this temperature dependence has to be determined by photoluminescence (PL) or optical absorption spectroscopy with caution. For example, it is at first reported that in InN the PL peak energy increases monotonically with increasing temperature\cite{T-InN-02}. A subsequent study yields normal redshift of optical absorption peak with increasing temperature, and attributes the anomalous blueshift of PL peak with increasing temperature to the fact that the PL peak is strongly affected by the localized states\cite{T-InN-03}. For semiconductor alloys, theoretical study on the temperature dependence of the band gap is rather at its preliminary stage. Attempt has been made to fit the experimental $E_g(x,T)$ based on the Varshni equation\cite{AlGaN-xT}. To our best knowledge, first-principles calculations on $E_g(T)$ for GaN, InN and $E_g(x,T)$ for their alloy Ga$_{1-x}$In$_x$N are still lacking. Of particular importance for the first-principles band structure calculations is the zero-point motion renormalization (ZPR), while the composition dependence of the Varshni paramete is crucial for the extrapolation of high-temperature $E_g(T)$ from the experimental low-temperature values. The present study aims at throwing some light on the application of VCA in the first-principles calculations of the temperature dependence of the band gap.
\section{Computational Method}
First-principles calculations are performed using the ABINIT program\cite{ABINIT-1}\cite{ABINIT-2}. In VCA, the potential of each atom in the alloy is replaced by a composition-weighted average of the potentials of its components. We employ the VCA provided by the ABINIT program. The pseudopotentials are generated using the fhi98PP code\cite{fhi98PP}. The Ga-3$d$ and In-4$d$ electrons are explicitly included in the valence. The electronic wave-functions are expanded in a plane-wave basis with a kinetic energy cutoff of 50 Hartree. We use a $6\times6\times4$ $k$-point mesh for Brillouin-zone sampling. Phonons are calculated using density-functional perturbation theory (DFPT)\cite{DFPT} on a $8\times8\times6$ $q$-point mesh. It is well known that the band gap calculated from LDA is severely underestimated. In the present study we apply the recently proposed TB09 functional\cite{TB09} (a modified version of the Becke-Johnson exchange\cite{BJ} part combined with a LDA PW92 correlation\cite{C-PW92} part). It allows us to improve the band gap prediction with a computational cost only slightly heavier than that of LDA. The electron-phonon matrix elements are calculated on a $8\times8\times6$ $q$-point grid in the adiabatic rigid-ion approximation\cite{EPI-Ponce}. An imaginary shift ($i\delta$) of 0.1 eV is used in the perturbation denominator of the sum-over-states to avoid numerical instability.
\newline
The most common empirical relation for the variation of the band gap with temperature in semiconductors is the so-called Varshni relation\cite{Varshni}:
\begin{equation}
E_g=E_0-\alpha T^2/(T+\beta)
\end{equation}
where $E_0$ is the band gap at 0 K and $\alpha$,$\beta$ are fitting parameters. Although the Varshni relation bears no physical resemblance and incorrectly predicts a quadratic temperature dependence for $T\rightarrow0$\cite{Cardona}, it fits the experimental data remarkably well. In present study, we continue to use the Varshni relation for its significant popularity in applications.
\section{Results and Discussions}
The calculated lattice constants and band gaps of GaN and InN with the TB09 functional are summarized in Table \uppercase\expandafter{\romannumeral1}. While LDA severely underestimates the band gap for GaN and even incorrectly predicts a metallic state for InN, the TB09 functional yields band gaps close to the experiments. Our VCA calculations show that the dependence of $a$ and $c$ on the In content $x$ clearly deviate from the Vegard's law\cite{Vegard}. In Fig. 1 we show the calculated and experimental band gaps. To study the effect of local environment relaxations (which VCA lacks) on the band gaps, we also perform calculations using the TB09 functional on fully relaxed SQSs. Firstly, the TB09 functional corrects for the LDA band gaps independently on $x$, which further confirms the applicability of LDA in calculating the band gap bowing parameter for Ga$_{1-x}$In$_x$N alloy\cite{GaInN-HSE06}\cite{GaInN-LDA12}. Secondly, our VCA calculations show that the band gap bowing at low In content is larger than that at high In content. Therefore one single composition-independent bowing parameter may not be adequate for accurate description of the nonlinear band gap bowing. If we enforce the composition-independent bowing, the corresponding bowing parameter reads 3.85 eV by a least-square fitting. Thirdly, the band gap bowing of SQS alloy is much smaller than that of VCA alloy. The tendency accidentally corresponds to the effect of In clustering where the band gap bowing ranges from 2.1 eV for the uniform case to 3.9 eV for the clustering case\cite{GaInN-LDA+C}. The calculated phonon density of states (DOS) is shown in Fig. 2. The peak of phonon DOS shifts to lower frequency while the shape of phonon DOS remains unchanged as In content $x$ is increased. Compared with 32-atom SQS calculations\cite{SQS-wz}, VCA succeeds in capturing the shift of phonon DOS, but fails in describing the broadening of the phonon spectra, since the broadening of phonon DOS upon alloying is associated with local environment disorder. 
\begin{table*}
\caption{\label{arttype} Lattice constants and band gaps of GaN and InN calculated with the TB09 functional. Experimental data are taken from Ref. \cite{property} and are shown in parentheses.}
%\begin{indented}
%\begin{ruledtabular}
\begin{tabular}{@{}cccc}
\hline
\br
&$a$(\AA)&$c$(\AA)&$E_g$(eV)\\
\br
GaN&3.216 (3.189)&5.239 (5.185)&2.93 (3.51)\\
InN&3.521 (3.545)&5.692 (5.703)&0.57 (0.78)\\
\br
\hline
\end{tabular}
%\end{ruledtabular}
%\end{indented}
\end{table*}
%\begin{figure}[htb!]
%\centering
%\includegraphics[scale=0.3]{lattice.eps}
%\caption{Lattice constants of Ga$_{1-x}$In$_x$N.}
%\end{figure}
\begin{figure}[htb!]
\centering
\includegraphics[scale=0.3]{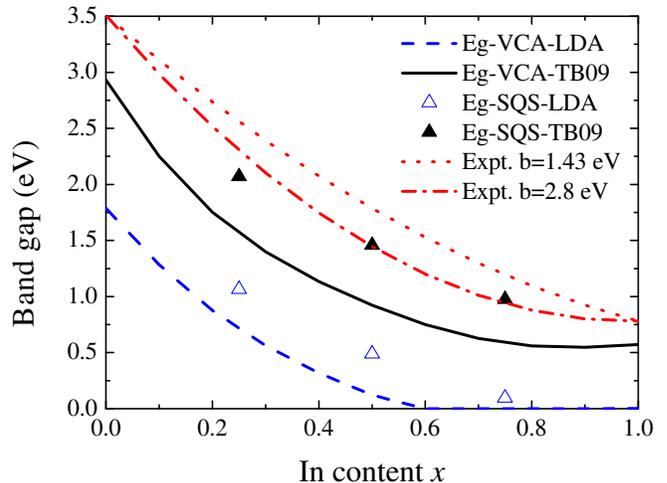}
\caption{Band gaps of Ga$_{1-x}$In$_x$N. Zero band gaps represent the incorrect metallic states predicted by LDA. SQSs are taken from Ref. \cite{SQS-wz}. Experimental bowing parameters are taken from Ref. \cite{GaInN-exp-1.43} ($b=1.43$ eV) and Ref. \cite{GaInN-exp-2.8} ($b=2.8$ eV).}
\end{figure}
\newline
\begin{figure}[htb!]
\centering
\includegraphics[scale=0.3]{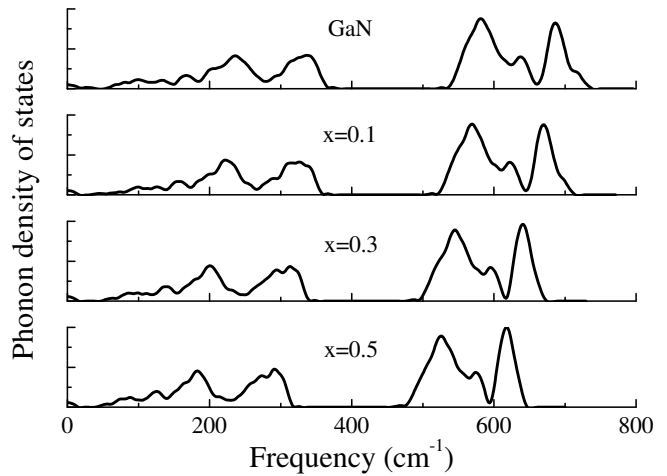}
\caption{Phonon density of states of Ga$_{1-x}$In$_x$N.}
\end{figure}
\newline
The temperature dependence of the band gap $\Delta E_g$(T) can be decomposed into two parts: the electron-phonon interaction (EPI) term and the thermal expansion (TE) term. While the EPI contribution to the temperature dependence of the band gap of GaN is calculated from first-principles, the TE contribution\cite{Cardona} is calculated from $-B(\frac{\partial E_g}{\partial p})_T\int_0^T[2\alpha_a(T')+\alpha_c(T')]dT'$. The bulk modulus $B$ is taken from $ab$ $initio$ calculatoins\cite{bulk_modulus}, while the pressure coefficient of band gap $(\frac{\partial E_g}{\partial p})_T$ is taken from low-temperature PL measurement\cite{pressure_coefficient}. The linear thermal expansion coefficient $\alpha(T)$ over the entire temperature range is described by the experimentally determined Debye model\cite{TEC}. Our calculations agree well with the experimental data\cite{T-GaN}, as is shown in Fig. 3. The ZPR is as large as -0.15 eV. The main contribution to the temperature dependence of the band gap comes from the EPI term, which is almost three times of the TE term. The fitted Varshni parameters (for only the EPI term) $\alpha$ and $\beta$ reads 0.51 meV/K and 745 K respectively. A simple average of the diverse experimental data suggests $\alpha$=0.91 meV/K and $\beta$=830 K\cite{35}, while a more recent measurement yields $\alpha$ in the range of 0.54-0.63 meV/K and $\beta$ in the range of 700-745K.
\begin{figure}[htb!]
\centering
\includegraphics[scale=0.3]{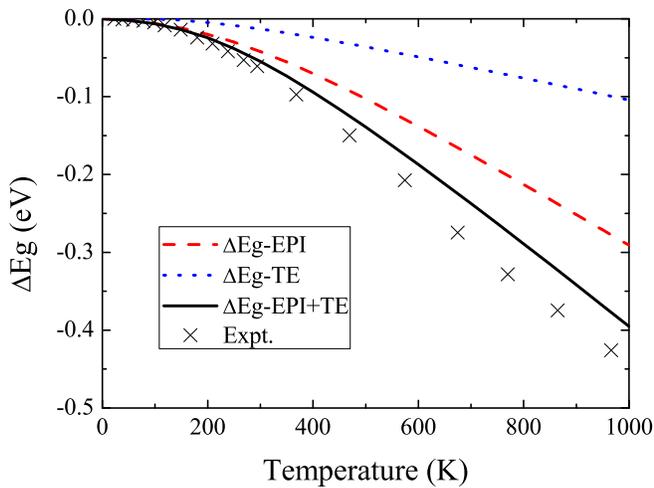}
\caption{Temperature dependence of the band gap (referenced to zero) of GaN. Experimental data are taken from Ref. \cite{T-GaN}.}
\end{figure}
\newline
The calculated $(x,T)$ dependence of the band gap is shown in Fig. 4. Due to the LDA band gap problem (see Fig.1), we focus on the Ga-rich ($x\le$0.5) contents. The ZPR and the Varshni parameter $\alpha$ decrease in magnitude as $x$ is increased because of the larger atomic mass of In, indicating a decrease in the strength of EPI. We describe the deviation of the composition dependence from linearity by a bowing term $-bx(1-x)$, as is shown in Fig. 5. The bowing parameter for the ZPR is $-$0.1 eV, which should be taken into account in the calculated band gap bowing parameter (in present study 3.85 eV). The extrapolated ZPR for InN is -0.036 eV, considerably smaller than that for GaN. The extrapolated $\alpha$ for InN is 0.096 meV/K, while the experimentally measured value reads 0.414 meV/K\cite{T-InN-03}. The large difference indicates a significant role of TE in the $E_g(T)$ of InN, contrary to the situation for GaN where EPI dominates over TE.
\begin{figure}[htb!]
\centering
\includegraphics[scale=0.3]{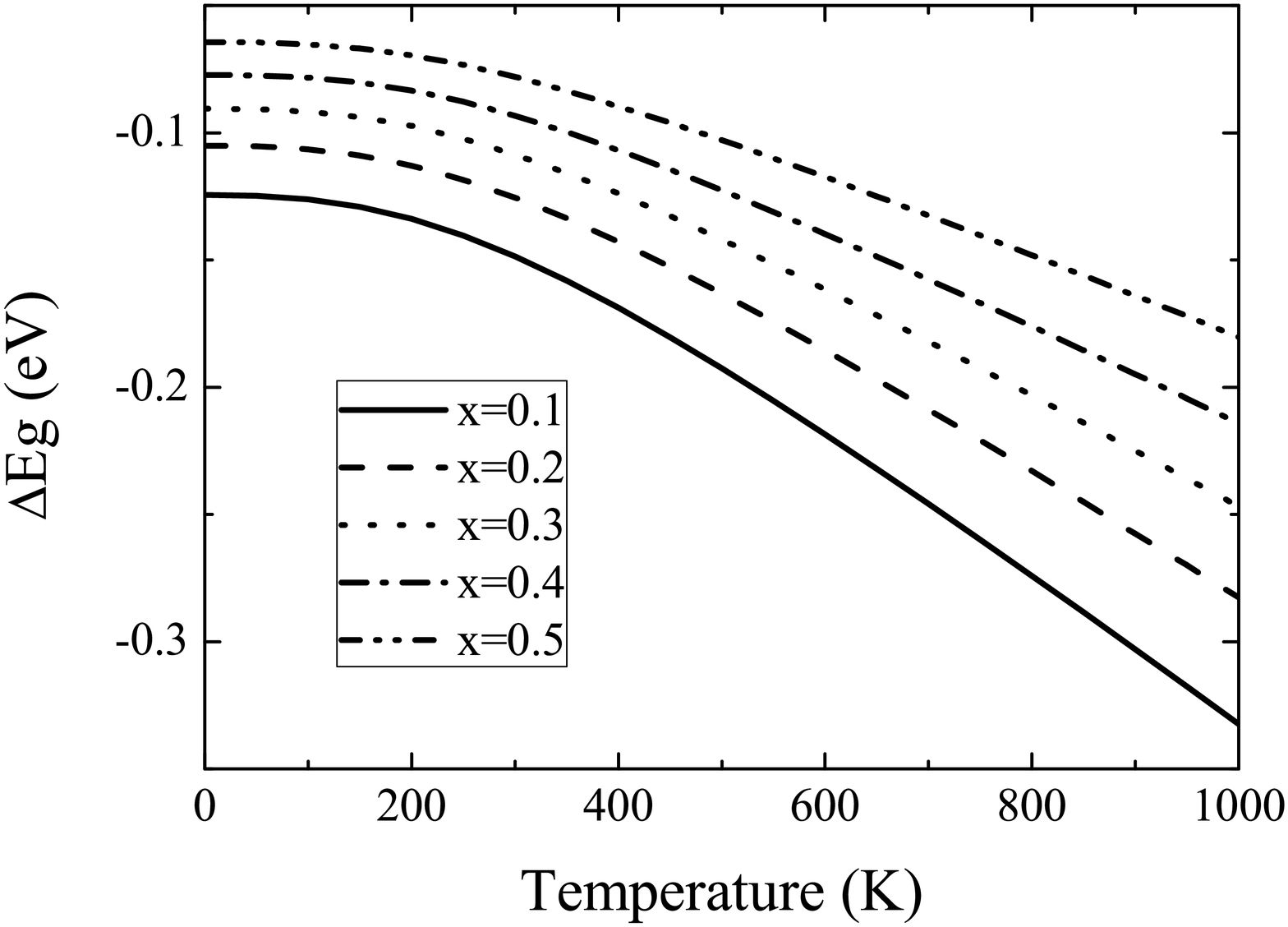}
\caption{Temperature dependence of the band gaps of Ga$_{1-x}$In$_x$N alloy at $x\le$0.5.}
\end{figure}
\begin{figure}[htb!]
\centering
\includegraphics[scale=0.3]{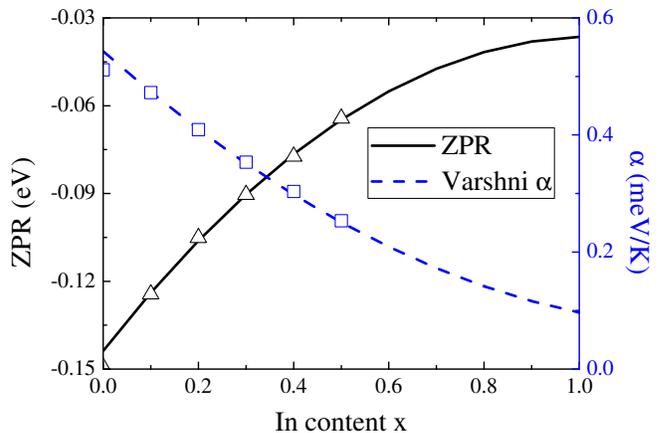}
\caption{Composition dependence of the ZPR and the Varshni $\alpha$.}
\end{figure}
\section{Conclusions}
In conclusion, we have studied the structural, electronic and vibrational properties of the Ga$_{1-x}$In$_x$N alloy using VCA from first-principles. We obtain a band gap bowing parameter of 3.85 eV with the TB09 functional. VCA succeeds in capturing the shift of phonon DOS, but fails in describing the broadening of the phonon spectra. We have also studied the role of EPI in the temperature dependence of the band gaps for GaN, InN and their alloy Ga$_{1-x}$In$_x$N. For GaN EPI plays the dominant role, while for InN TE could contribute significantly. The calculated ZPR is important for modifying theoretical zero-temperature band gap, while the fitted Varshni parameter $\alpha$ is crucial for extrapolating high-temperature band gap experimentally.

\ack
This research used computational resources at the Center for Functional Nanomaterials, Brookhaven National Laboratory, which is supported by the US Department of Energy under Contract No. DE-AC02-98CH10886. Work at Stony Brook was supported by US DOE Grant No. DE-FG02-08ER46550 and DE-FG02-09ER16052. Jian Liu is also sponsored by the China Scholarship Council (CSC).

\section*{References}
\bibliographystyle{iopart-num}
\bibliography{ref}

%\end{thebibliography}

\end{document}